\documentclass[journal=ascefj,manuscript=letter]{achemso}
\setkeys{acs}{articletitle=true, keywords=true, doi=true}

\usepackage{chemformula} 
\usepackage[T1]{fontenc} 
\usepackage{amssymb,amsmath}
\usepackage{mathtools}
\usepackage{hyperref}

\SectionsOn

\author{Damien Drix}
\email{d.drix@herts.ac.uk}
\author{Michael Schmuker}
\affiliation[UHerts]{Biocomputation group, Department of Computer
Science, University of Hertfordshire, Hatfield, United Kingdom}

\title{Resolving fast gas transients with Metal-Oxide sensors}

\begin{document}


\begin{abstract}
Electronic olfaction can help detect and localise harmful gases and
pollutants, but the turbulence of natural environment presents a
particular challenge: odor encounters are intermittent, and an effective
electronic nose must therefore be able to resolve short odor pulses. The
slow responses of the widely-used Metal-Oxide (MOX) gas sensors
complicate the task. Here we combine high-resolution data acquisition
with a processing method based on Kalman filtering and absolute-deadband
sampling to extract fast onset events. We find that our system can
resolve the onset time of odour encounters with enough precision for
source direction estimation with a pair of MOX sensors in a stereo-osmic
configuration.
\end{abstract}


\hypertarget{introduction}{%
\section{Introduction}\label{introduction}}

Electronic olfaction has potential in many areas such as industrial and
environmental monitoring and safety, where it can help detect and
localise harmful gases or pollutants.

But in natural environments, odors are dispersed by turbulent plumes and
encounters are intermittent \citep{Mylne:1991da}. The temporal
statistics of these odor pulses (hereafter also called \emph{bouts})
contain information about source location \citep{Schmuker:2016hq}. An
effective electronic nose thus needs to resolve both short pulses and
pulses in rapid succession. Metal-Oxide (MOX) gas sensors are widely
used, but have impulse response durations in the order of tens to
hundreds seconds \citep{Pashami:2012jt}, and are therefore often thought
to be of limited utility in turbulent environments.

However, a large part of the impulse response is due to a slow sensor
recovery phase, in the order of 100s, in which the sensor conductance
slowly returns to baseline while the initial reaction of the volatile
with the sensor electrode is reversed \citep{Korotcenkov:2008dh}. The
onset of the response itself is near-instantaneous and can be detected
after fractions of a second. Therefore, repeated short-duration bouts
could be detectable with the help of specific physical mitigation (for
instance sensor purging \citep{Gonzalez:2011cz} or pulsed heating
\citep{Vergara:2014gp}) or through signal processing that separates the
initial binding from the recovery phase.

Here we built a multi-channel MOX sensor electronic nose with high bit
depth and sampling rate to investigate how much can be achieved through
signal processing alone. We developed a signal processing method based
on a Kalman filter and absolute deadband sampling to isolate successive
bouts and encode their onset time. We demonstrate the system's ability
to resolve onset times and repeated bouts in a stereo-enose setup that
infers the direction of a puff of odorant from stereo delays.

\hypertarget{results}{%
\section{Results}\label{results}}

\hypertarget{data-acquisition}{%
\subsection{Data acquisition}\label{data-acquisition}}

Our gas sensor boards consist of four metal-oxide (MOX) sensors and a
high-resolution analog-to-digital converter (ADC). We use four sensors
manufactured by Figaro Inc.~(Osaka, Japan): TGS2600, TGS2602, TGS2610
and TGS2620, to cover a wide range of target gases. The ADC (ADS122C04,
Texas Instruments) offers 24 bits of resolution and can sample all four
sensor channels at a frequency of up to 200 Hz. As MOX sensors are
affected by ambient temperature and humidity, the boards can also host
an optional 16-bit digital temperature and humidity sensor (SHT31-DIS,
Sensirion).

\begin{figure}
\hypertarget{fig:schematic}{%
\centering
\includegraphics{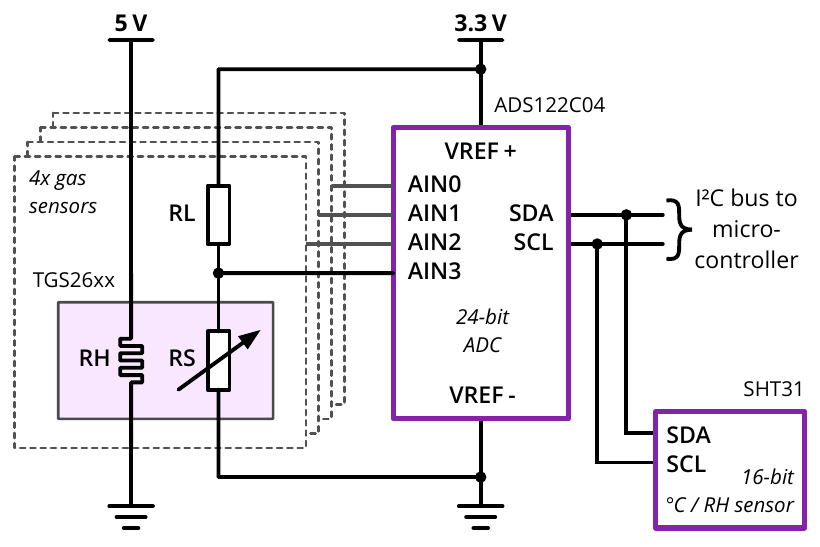}
\caption{\textbf{Simplified schematic of the sensor boards.} The ADC on
each sensor board measures the voltage across the MOX sensing elements
RS. A separate 5V supply powers the heating elements RH. The sensor
board communicates via I2C with a Cortex-M7 microcontroller that
transmits the data to the host computer.}\label{fig:schematic}
}
\end{figure}

An I2C bus operating at 800~KHz connects the boards to a microntroller
(Teensy 4.0, PJRC.COM) that reads out the data and transmits it to the
host computer via USB (fig.~\ref{fig:schematic}). The system is set up
so that the microcontroller can handle multiple sensors in parallel, for
instance left and right electronic noses in a stereo configuration.

\begin{figure}
\hypertarget{fig:curves}{%
\centering
\includegraphics{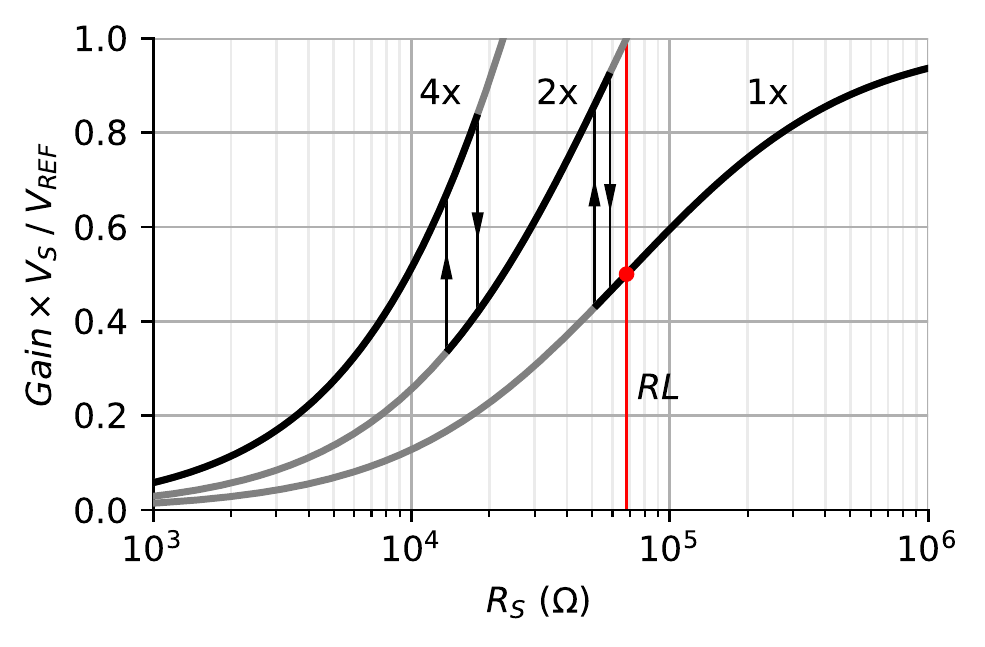}
\caption{\textbf{Automatic input gain selection maintains sensitivity
over a large input range.} ADC measurement for a varying sensor
resistance \(RS\) at gain settings 1x, 2x and 4x. \(RS\) follows an
approximate power law with respect to gas concentration; thus it makes
sense to use a logarithmic scale, where the slope of the measurement
function indicates the sensitivity to a relative change (eg.
\(\pm 1 \%\) in gas ppm). Thin arrows indicate the thresholds at which
we increase or decrease the input gain to avoid the regions of lower
sensitivity. The red line indicates the load resistance \(RL\) for
reference.}\label{fig:curves}
}
\end{figure}

The gas sensors are connected to the ADC in the voltage divider
configuration that is standard for this type of sensor
\citep{Arshak:2003kv}, with the sensing element RS in series with a load
resistor RL (fig.~\ref{fig:schematic}). This configuration works well
with the chosen ADC as it allows a ratiometric measurement relative to
the power supply, free from common-mode noise. On the other hand its
sensitivity degrades as RS deviates from RL, which normally requires an
adjustable RL calibrated for a specific sensor at a given temperature
and expected gas concentration. Here we make use of the ADC's high
resolution and variable input gain instead, which lets us pick a fixed
load resistor \(RL = 68\, \mathrm{k\Omega}\) and still maintain a good
sensitivity through a large range of concentrations and ambient
conditions (fig.~\ref{fig:curves}).

The ADC measures a ratio \(x = \frac{V_S}{V_S + V_L}\), where \(V_S\)
and \(V_L\) are the voltages across RS and RL, respectively. From this
we compute the relative conductance \(g_{rel}\) of each sensor relative
to its load resistor: \[
g_{rel} = \frac{g_S}{g_L} = \frac{V_L}{V_S} = \frac{1}{x} - 1
\]

Then, we divide by the baseline values at the start of each recording to
get the normalised sensor conductance \(g\) at time \(t\): \[
g(t) = \frac{g_{rel}(t)}{g_{rel}(0)}
\]

The purpose of the normalisation is to use the same parameters for
processing multiple sensors with different characteristics. It is not
fundamentally required, since none of the algorithms assume a specific
or constant baseline value for \(g\).

In preliminary work we had found the quantization noise from 10-bit ADCs
to be not insignificant compared to the signals of interest,
complicating downstream processing. The 24-bit ADC solves this problem
and its good noise performance lets us resolve very low-amplitude
fluctuations (see \emph{Supporting Information} fig.~S-1).

\hypertarget{experimental-setup}{%
\subsection{Experimental Setup}\label{experimental-setup}}

\begin{figure}
\hypertarget{fig:setup}{%
\centering
\includegraphics{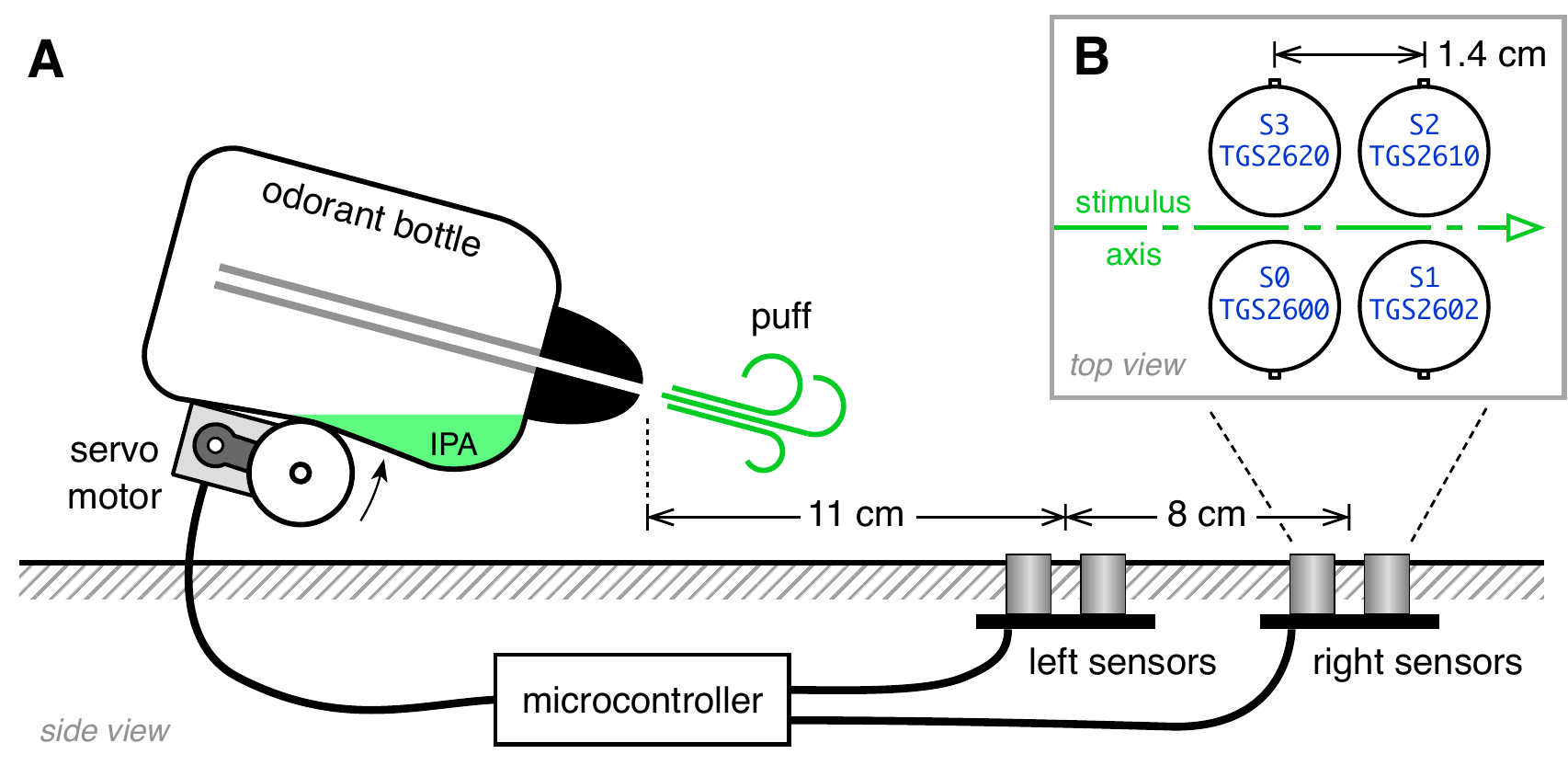}
\caption{\textbf{An automated setup delivers puffs of odorant towards
stereo sensor boards.} \textbf{A}: Side view of the system in its
left-to-right configuration. \textbf{B}, inset: top view of a sensor
board showing the position of the four sensors in relation to the
stimulus axis.}\label{fig:setup}
}
\end{figure}

We deliver puffs of isopropyl alcohol (IPA) to the sensors by means of a
soft plastic bottle (NeilMed Inc, USA) squeezed by a servomotor to force
vapours out of the nozzle (fig.~\ref{fig:setup}). This creates a sharp
puff that we could observe up to 50 cm from the nozzle in a quiet
atmosphere.

We record simultaneously from two identical boards placed along the
direction of travel of the puff. This yields four pairs of stereo
channels (S0 to S3), one for each MOX sensor type. We aim the stimuli
slightly downwards onto a flat surface and position the top of the MOX
sensors flush with that surface to reduce turbulence caused by the
sensors themselves, which might otherwise disrupt the narrow odor plume
before it reaches sensors on the far side. We record one dataset with
the stimulus traveling in the left-to-right direction, then move the
bottle to the other side and record another dataset for the
right-to-left direction.

\hypertarget{post-processing}{%
\subsection{Post-processing}\label{post-processing}}

\begin{figure}
\hypertarget{fig:kalman}{%
\centering
\includegraphics{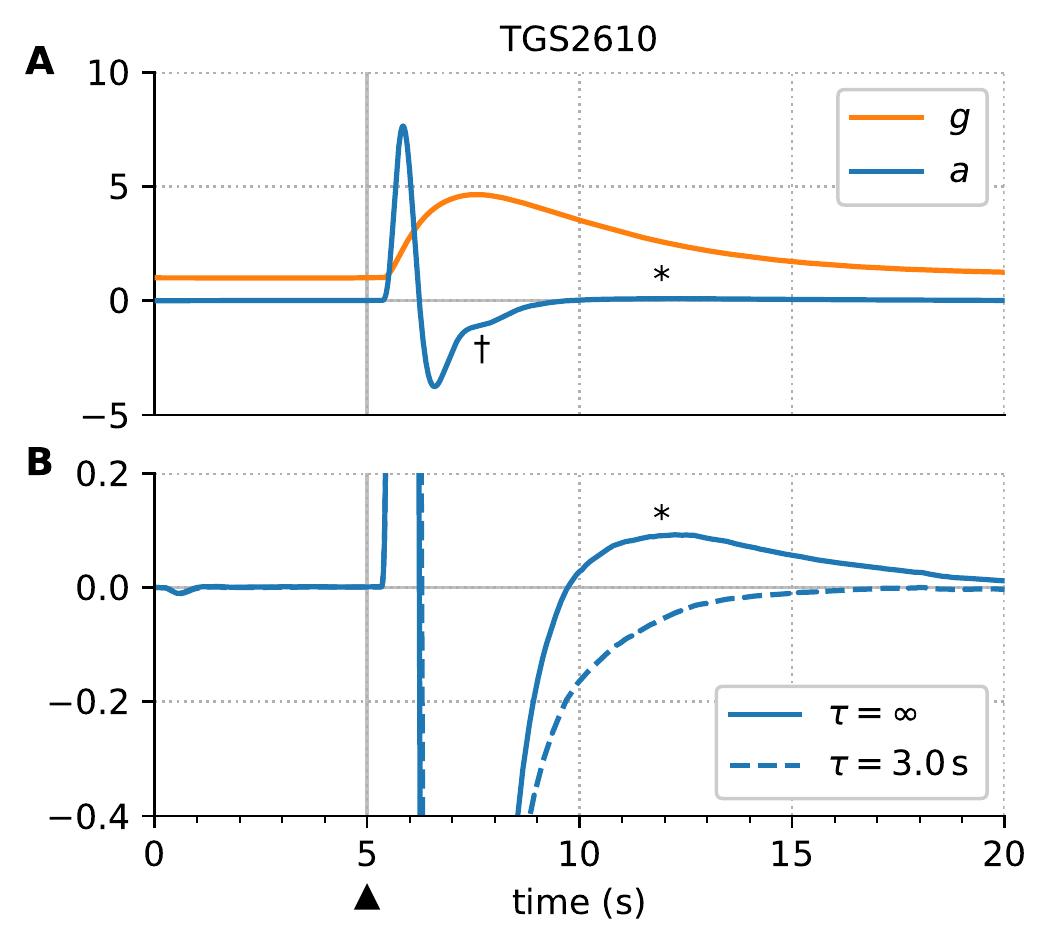}
\caption{\textbf{Kalman filtering recovers the onset of odorant bouts.}
The arrowhead (\(\blacktriangle\)) marks the time when the odorant
bottle is squeezed. \textbf{A}: conductance \(g\) and its second
derivative \(a\) estimated by the Kalman filter. \textbf{B}: Zooming in
shows the effect of the filter parameter \(\tau\) on the late-phase
response. The second peak (\(\ast\)) is effectively removed with
\(\tau = 3\) seconds without affecting the early response. The feature
marked \(\dagger\) is unrelated and probably caused by a transient
disturbance of the sensor.}\label{fig:kalman}
}
\end{figure}

\begin{figure}
\hypertarget{fig:spikes}{%
\centering
\includegraphics{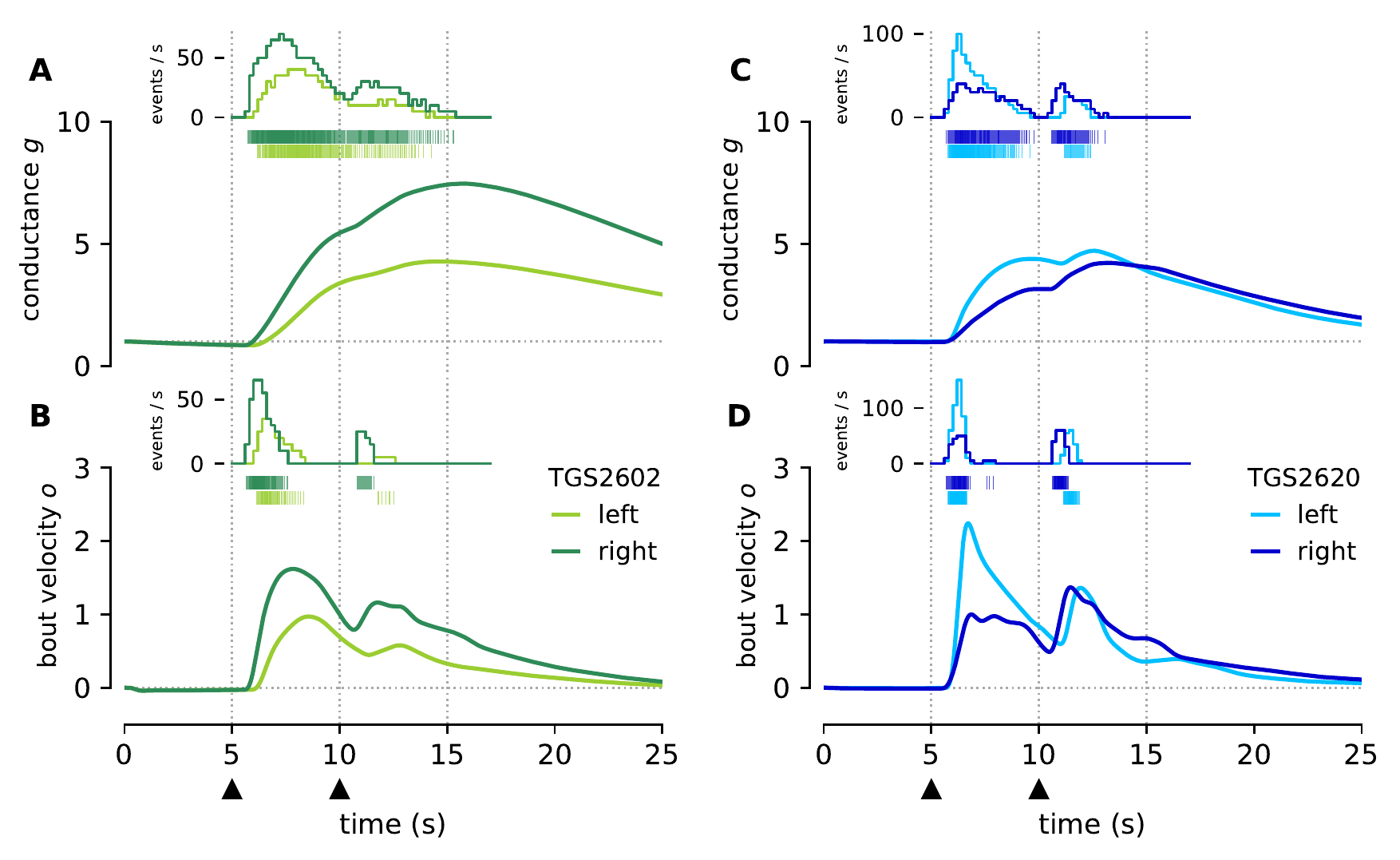}
\caption{\textbf{Events generated from the bout velocity variable
isolate the onset of each bout.} Responses of two sensor pairs during
the same trial with two puffs of odorant (\(\blacktriangle\)) at a
5-second interval in the right-to-left direction. \textbf{A, B}:
conductance \(g\) and bout velocity \(o\) estimated by the Kalman filter
for the TGS2602 sensors (left \& right), together with the resulting
events (thin vertical lines) and event rate (time histogram). \textbf{C,
D}: same as A \& B, but with the TGS2620 sensors, which have a faster
response time.}\label{fig:spikes}
}
\end{figure}

MOX sensors respond to puff of odorants with a fast rising phase,
followed by a slower decay back to baseline (fig.~\ref{fig:kalman} A).
This slow decay can mask fast transients, for instance when two bouts
occur close together in time (see e.g.~fig.~\ref{fig:spikes} A). The
goal of post-processing is to isolate the onset of the rising phase,
thus providing the ability to resolve short odor pulses. Various
solutions have been explored in previous work, such as taking the second
derivative or deconvolution based on an estimate of the sensor's impulse
response function \citep{Schmuker:2016hq}, blind deconvolution
\citep{Martinez:2019bf}, and band-pass filters \citep{Burgues:2019dz}.

Here we use a constant-acceleration Kalman filter to compute a denoised
estimate of the second derivative of the signal. The second derivative
peaks at the onset of each puff (fig.~\ref{fig:kalman} A). However it
also has a second, smaller positive peak when the relaxation slows down,
since that registers as a positive acceleration (fig.~\ref{fig:kalman}
B). As this could cause spurious bout detections (it is very small
compared to the onset response, but still 5 times larger than our bout
detection threshold) we modify the filter to suppress the second peak.
We do this by incorporating an exponential decay term into the system
equations for the first derivative \(v\), thus removing the expected
relaxation from the residual second derivative \(a\): \[
\begin{split}
g(t+dt) & = g(t) + v(t)\, dt + \left(a(t) - \frac{v(t)}{\tau}\right) \frac{dt^2}{2} \\
v(t+dt) & = v(t) + \left(a(t) - \frac{v(t)}{\tau}\right) dt \\
a(t+dt) & = a(t)
\end{split}
\]

The parameter \(\tau\) sets the time constant of the decay term. We
estimate it empirically for each sensor type, selecting the highest
value which still suppresses the second peak. We also define a variable
\(o = \int a(t)\, dt\), which we call \emph{bout velocity}. Being the
integral of the residual second derivative \(a\), this variable is
essentially a first derivative of the signal, like \(v\), but with the
second peak removed.

\hypertarget{event-based-onset-encoding}{%
\subsection{Event-based onset
encoding}\label{event-based-onset-encoding}}

For the purpose of bout detection it is the time of the odorant onset
that matters, more than the precise time course of the sensor
conductance during and after the bout. Therefore it makes sense to
transform the continuous filter output into an event-based
representation that only transmits information during periods of
increasing conductance.

We employ an event-based encoding related to delta modulation that has
been variously called \emph{deadband sampling}
\citep{Vasyutynskyy:2007cr} or \emph{send-on-delta}
\citep{Miskowicz:2006cq}, and is also used in the DVS camera
\citep{Lichtsteiner:2008bm}.

If the variable's value at a time \(t\) is \(z(t)\) and the time of the
last event is \(t_{prev}\), then a new event is emitted whenever the
difference exceeds a certain threshold \(\theta=0.02\): \[
|z(t) - z(t_{prev})| > \theta
\] This form of absolute deadband sampling yields a stream of events
with an instantaneous rate \(f(t) \propto |\frac{d}{dt} z(t)|\), as the
algorithm differentiates. The sign of the difference indicates whether
the variable increased (ON events) or decreased (OFF events). We discard
all OFF events as these do not correspond to the onset of a bout.

This event-based encoding can be applied to the sensor conductance
(\(z = g\), for an event rate \(f\) proportional to \(v\)) as well as to
the bout velocity variable (\(z = o\), for an event rate \(f\)
proportional to the filter output \(a\)). We find that when applied to
the filter output, it produces well-separated bursts of events for two
bouts separated by 5 seconds, a much shorter delay than the recovery
phase of the sensor conductance (fig.~\ref{fig:spikes} B, D). On the
other hand, events generated from the sensor conductance are prone to
merging into a single burst when bouts follow each other closely
(fig.~\ref{fig:spikes} A).

\hypertarget{direction-detection-in-stereo-osmic-configuration}{%
\subsection{Direction Detection in Stereo-osmic
Configuration}\label{direction-detection-in-stereo-osmic-configuration}}

\begin{figure}
\hypertarget{fig:delays}{%
\centering
\includegraphics{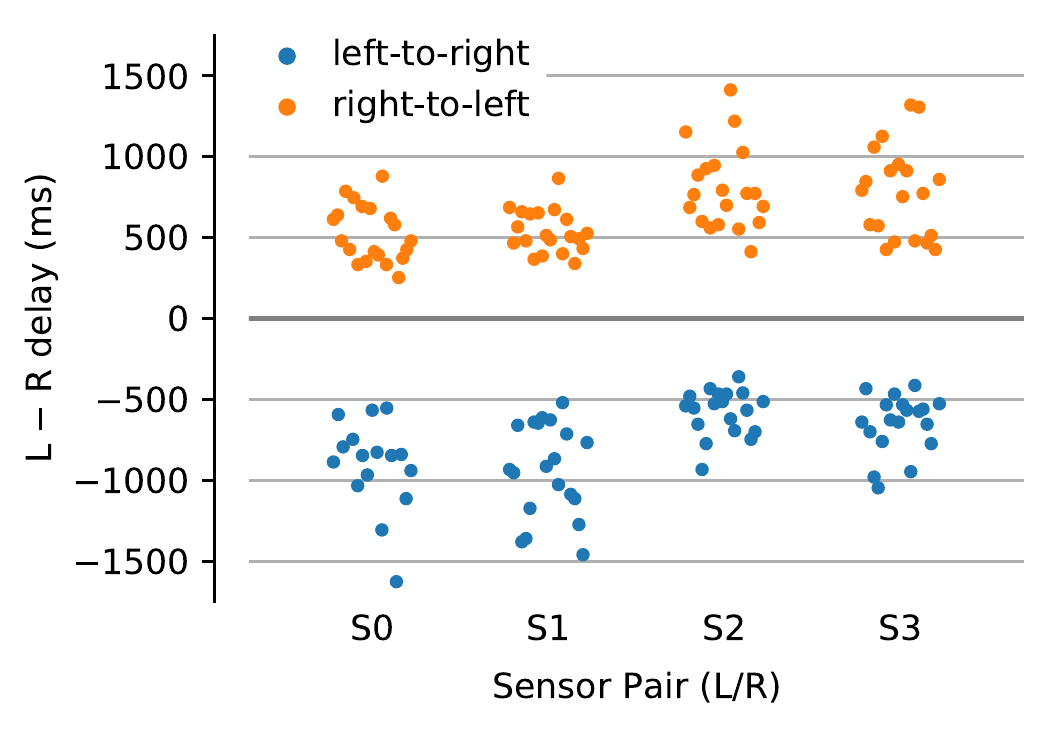}
\caption{\textbf{Relative delays between left and right channels encode
the direction of travel}. Shown here are the delays between the first
event on the left channel and the first event on the right channel over
40 trial runs, colour-coded by stimulus direction (20 trials with a
left-to-right puff and 20 with a right-to-left puff). Three outliers
with a delay greater than 2 seconds are not shown on this
graph.}\label{fig:delays}
}
\end{figure}

We apply this event-based encoding to the data obtained from our
recording setup in stereo-osmic configuration (fig.~\ref{fig:setup}). As
the stimulus travels over the sensors, the left and right sensor boards
will detect its onset at slightly different times, with the delay
between left and right boards depending on the speed and direction of
the puff. We extract the time of the first event on each channel, and
then compute the time differences between the left and right sensors
(fig.~\ref{fig:delays}). We find that the sign of that time difference
encodes the direction of the stimulus unambiguously, despite some
variance due to turbulent flow. A slight systematic offset is apparent
between channels. We have observed a similar effect when the axis of the
puff deviates from the center line; thus the offset may be due to
lateral flow, although mismatched sensor characteristics may also play a
role.

\hypertarget{discussion}{%
\section{Discussion}\label{discussion}}

We show that off-the-shelf metal-oxide sensors can resolve relative
onset delays in the sub-second range using a Kalman filter to extract
bout onset times. This software approach could be combined with hardware
measures \citep{Gonzalez:2011cz, Vergara:2014gp} to increase the
temporal resolution of the data even further.

The proposed method is inherently robust to baseline drift, a common
issue with MOX sensors where their conductance at a certain odorant
concentration will vary over the lifetime of the sensor and across the
slower changes in ambient temperature and humidity.

We find that bout onset information lends itself well to event-based
encoding and processing. In the past decade, event-based sensors have
garnered interest because they are efficient in bandwidth (transmitting
only changes rather than redundant frames) and make certain tasks
simpler. While electronic olfaction is still low-bandwidth and
low-dimensional compared to vision, this may change as sensor technology
improves. Inferring stereo delays from event timing is less
computationally expensive than, for instance, from the cross-correlogram
of the two channels.

In our experiment, the timing of bout onset events is enough to estimate
the direction of a puff of odorant, just like bout frequency was already
known to contain information about the the distance to the source
\citep{Schmuker:2016hq}. Future research should confirm how well that
event-based approach translates to real-world conditions, and assess to
which extent event timing is sufficient to navigate towards an odor
source with more chaotic plumes and lower odorant concentrations.

For instance, the ability of the system to resolve successive bouts at
short intervals should be explored systematically. From the data in
fig.~\ref{fig:spikes} we estimate that we can separate successive bouts
down to an interval of about 1 to 3 seconds, but this would have to be
quantified in conjunction with ground truth data about the plume
structure.

Finally, while we have not quantified the effect of bit depth on the
stereo detection task presented here (which uses relatively high
concentrations), our system's ability to resolve very low-amplitude
fluctuations may also be advantageous when applied to low-concentration
plumes.

\hypertarget{conclusion}{%
\section{Conclusion}\label{conclusion}}

Our work demonstrates how a relatively simple and lightweight setup
using MOX sensors can extract onset times and separate successive bouts
at intervals much shorter than the sensor's recovery time. Increased
temporal resolution renders metal-oxide sensors more useful in
extracting data from turbulent plumes, in particular when the temporal
structure of gas concentration fluctuations is of concern, rather than
absolute concentration values.

This highlights their potential for odor-guided navigation in embedded
systems such as weight-constrained aerial vehicles
\citep{Burgues:2019gf}. In ground-based robots, the proposed method
could remove the need for a separate anemometer to estimate odor source
direction.

While the present work focused on temporal resolution, future research
should assess whether particular odorants and mixtures of odorants can
be reliably identified using the event-based approach. This would
constitute a purely event-based system for the simultaneous
identification and localisation of gas sources.

\hypertarget{supporting-information}{%
\section{Supporting Information}\label{supporting-information}}

\setcounter{figure}{0}    
\let\origfigure=\figure
\let\endorigfigure=\endfigure
\renewenvironment{figure}[1][]{%
  \origfigure[H]
}{%
  \endorigfigure
}
\renewcommand{\thefigure}{S-\arabic{figure}}

\begin{figure}
\hypertarget{fig:bitdepth}{%
\centering
\includegraphics{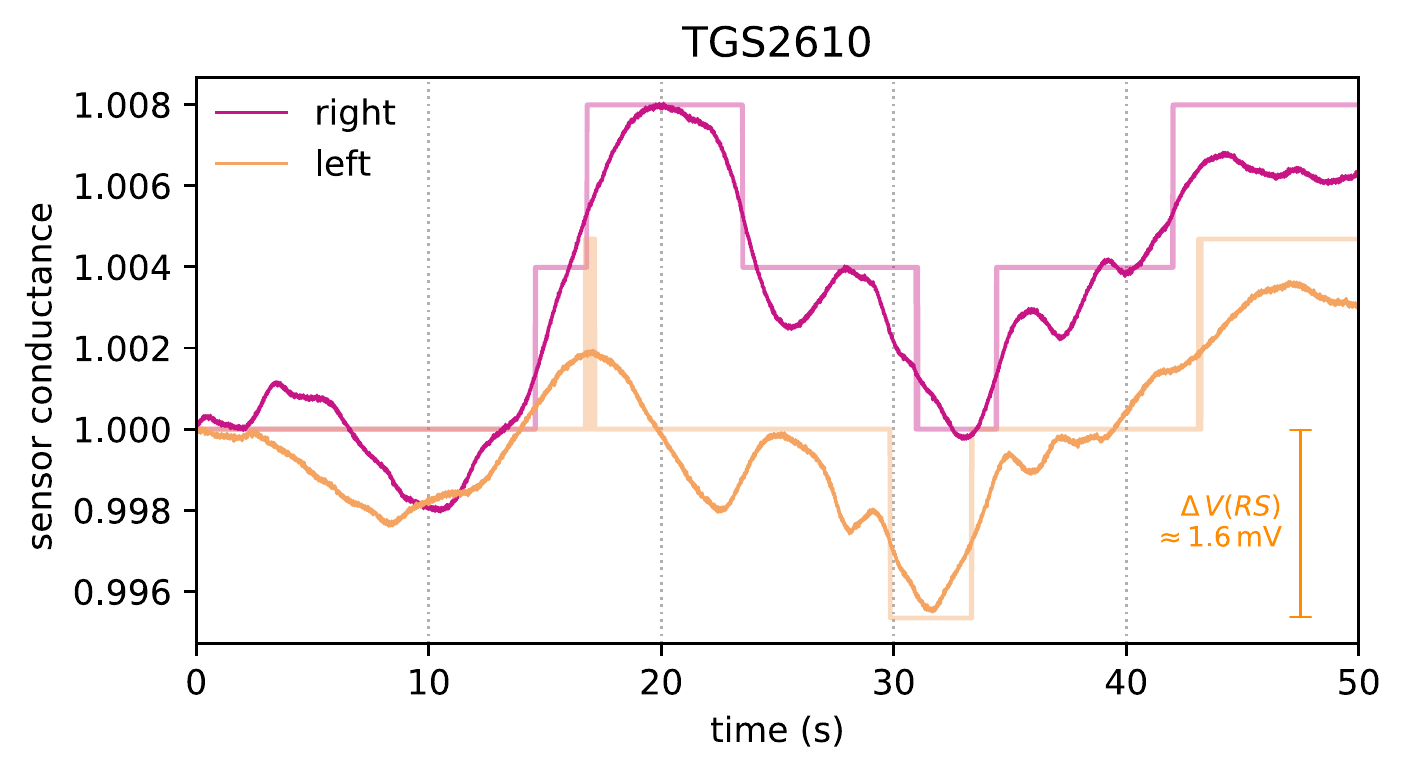}
\caption{\textbf{High-resolution acquisition resolves background
turbulence.} Data acquired from a stereo pair of MOX sensors in
uncontrolled environmental conditions (university office) shows
low-amplitude background fluctuations with temporal correlations between
left and right channels. Dark lines: normalised sensor conductance \(g\)
acquired by the 24-bit ADC (23 bits for positive values). Pale steps:
simulated 12-bit ADC (11 bits for positive values), generated by setting
the 12 lower bits of the 23-bit data to zero. The scale of the
corresponding ADC input voltage is shown for the left sensor (orange),
assuming a \(+3.3 \, \mathrm{V}\) full-scale range. The height of the
scale bar corresponds to a least-significant bit (LSB) size of
\(1.6 \, \mathrm{mV}\) for 11-bit data. LSB size for 23-bit data is
approximately \(0.4 \, \mathrm{\mu V}\). A scale bar for the right
sensor would be slightly different because each channel is normalised
separately.}\label{fig:bitdepth}
}
\end{figure}


\begin{acknowledgement}
DD and MS were funded from EU H2020 Grants 785907 and 945539 (Human
Brain Project SGA2 and SGA3 ). MS was funded by MRC grant MR/T046759/1
(NeuroNex: From Odor to Action).
\end{acknowledgement}


\bibliography{manuscript}

\end{document}